\documentclass[aps,twocolumn,epsfig]{revtex4}

\usepackage{amssymb}
\usepackage{amstext}
\usepackage{amscd}
\usepackage{amsthm}
\usepackage{amsopn}
\usepackage{amsmath}
\usepackage{amsbsy}
\usepackage{indentfirst}
\usepackage[dvips]{graphicx}
\usepackage{mathrsfs}
\usepackage{color}
\usepackage{xcolor}

\usepackage{ulem}
\definecolor{DarkRed}{rgb}{0.5,0.0}

\newcommand{\Cal}[1]{{\cal #1}}
\newcommand{\be}{\begin{equation}}
\newcommand{\ee}{\end{equation}}
\newcommand{\bea}{\begin{eqnarray}}
\newcommand{\eea}{\end{eqnarray}}
\newcommand{\p}{\partial}

\newcommand{\ind}[1]{\begin{footnotesize}\mbox{#1}\end{footnotesize}}

\newcommand{\VM}[1]{\big\langle #1 \big\rangle}

\newcommand{\f}{\textsc{f}}

\newcommand{\XXX}[1]{}

\newcommand{\Hin}{\mathscr{H}_{\ind{in}}}
\newcommand{\Hout}{\mathscr{H}_{\ind{out}}}
\newcommand{\inc}{{\ind{in}}}
\newcommand{\out}{{\ind{out}}}
\newcommand{\psg}{\Cal P}
\newcommand{\qsg}{\Cal Q}
\newcommand{\V}{V_{\textsc{SD}}}
\newcommand{\TB}{T_{\textsc{B}}}
\newcommand{\thetaB}{\theta_{\textsc{B}}}
\newcommand{\Ibs}{I_\textsc{bs}}
\newcommand{\Fbs}{F_\textsc{bs}}

\renewcommand{\Re}{{\mathrm{Re}}}
\usepackage{tabularx}
\newcolumntype{C}[1]{>{\centering\arraybackslash}p{#1}} % zentrierte Spalten mit Breitenangabe 
\newcolumntype{R}[1]{>{\raggedleft\arraybackslash}p{#1}} % rechtsb?ndig mit Breitenangabe 

\newcommand{\Ci}{\mathrm{i}} 	% Komplex i
\newcommand{\e}{\mathrm{e}} 	% e-Funktion
\newcommand{\diff}{\mathrm{d}}

\newcommand{\bra}[1]{\langle #1 \vert}
\newcommand{\ket}[1]{\vert #1 \rangle}

\newcommand{\erwartungswert}[1]{\langle #1 \rangle}

\newcommand{\op}{\hat}

\newcommand{\opcd}{\op c^\dagger}
\newcommand{\opcp}{\op c^{\phantom{\dagger}}}

\newcommand{\hc}{\text{H.c.}}

\newcommand{\R}[1]{\mathrm{#1}}

\begin{document}

\author{A. Bransch{\"a}del}
\affiliation{Institut f\"ur Theorie der Kondensierten Materie, Karlsruhe Institute of Technology, 76021 Karlsruhe, Germany}
\author{E. Boulat}
\affiliation{Laboratoire MPQ, CNRS UMR 7162, Universit\'e Paris Diderot, 75205 Paris Cedex 13}
\author{H. Saleur}
\affiliation{Institut de Physique Th\'eorique, CEA, IPhT and CNRS, URA2306, Gif Sur Yvette, F-91191}
\affiliation{Department of Physics, University of Southern California, Los Angeles, CA 90089-0484}
\author{P. Schmitteckert}
\affiliation{Institute of Nanotechnology, Karlsruhe Institute of Technology, 76344 Eggenstein-Leopoldshafen, Germany }

\begin{abstract}
By using two independent and complementary approaches, we compute exactly the shot noise in an out-of-equilibrium interacting impurity model, the Interacting Resonant Level model at its self-dual point. An analytical approach based on the Thermodynamical
Bethe Ansatz allows to obtain the density matrix in the presence of a bias voltage, which in turn
allows for the computation of any observable. A time-dependent Density Matrix Renormalization Group
technique, that has proven to yield the correct result for a free model (the Resonant Level Model)
is shown to be in perfect agreement with the former method.
\end{abstract}

\title{Shot noise in the self-dual Interacting Resonant Level Model}
\maketitle

The study of transport in out of equilibrium, strongly interacting systems, is  of crucial importance for  technological applications. It also addresses some of the most challenging fundamental questions,  from the possibility of fluctuations theorems out of equilibrium \cite{Esposito08}, to the time evolution of  quantum many body entanglement  \cite{Klich09}. 

While experimental progress in this area has been swift and steady (see e.g. \cite{Reznikov95,Kumar96,Reulet03,Bomze05,Gustavsson06}), the theory has been held back by considerable technical difficulties. The physics of interest occurs usually in non perturbative regimes, where analytical methods are few and limited. Numerical approaches require real time simulations, which up to recently had been notoriously difficult.

Yet the situation is changing, especially in the case of quantum impurity problems.  Extensions of the Bethe ansatz to study transport properties  have been proposed \cite{Fendley96,Mehta06}, while time dependent DMRG (td-DMRG) calculations have become quite reliable. This led to the breakthrough in  \cite{Boulat08} where an exact, extremely non linear I-V characteristic was obtained analytically for the interacting resonant level model (IRLM) in  the scaling limit, and recovered with remarkable accuracy in td-DMRG simulations on a lattice model. The results can now be used as a benchmark for approximate approaches.

Interesting as it may be, an I-V characteristic does not capture the whole physics of a problem. In fact, the theoretical and experimental activity has concentrated lately on fluctuations,  embodied at vanishing temperature by the shot noise. Current fluctuations can be related to the charge of the ``elementary" quasiparticles, and their study is akin to  a subtle ``many-body spectroscopy".

While the free case has been studied in details  \cite{Levitov93,Levitov96,Klich09}, progress on the interacting case has been very sparse. 
Tackling the shot noise certainly increases the level of difficulty. On the analytical side, while the community has slowly accepted the idea that I-V characteristics could be obtained using the Bethe ansatz (thanks in part to alternate derivations, using all orders perturbation theory \cite{FLeS95}, or mappings onto an equilibrium system 
\cite{BLZ}), there is, to our knowledge,  only one published exact shot noise calculation \cite{FLS95}, which has not yet been confirmed. On the numerical side, extracting the noise from td-DMRG requires looking into the power spectrum of two point correlation functions, a quantity found to be strongly affected by finite size effects. 

Nevertheless, it has been realized very recently that the shot noise could indeed be reliably extracted from real time simulations, the finite size effects being controllable partly by analytical arguments, 
%{and partly by using some damping boundary conditions} %% PS: actually, those didn't work too well, did they?
and partly by linear finite size extrapolations \cite{previous}. 
We use this idea to study the noise in the  IRLM at the self dual point. 
Extending and adapting the ideas in \cite{FLS95} we are able to obtain this noise analytically in the scaling limit, 
while  we are also able to carry out the proposal of \cite{previous} to determine this noise numerically. 
These two are  in excellent agreement.

The IRLM describes a single fermionic level ($d^\dagger$) 
%with onsite energy tuned by a gate $\epsilon_d$, 
that is coupled to two leads of spinless electrons playing the role of thermodynamical baths (reservoirs of charge and energy). The coupling involves a tunneling term that allows electrons to hop from the leads to the impurity, and an impurity/lead capacitive coupling (coulombic repulsion $U$).
After the usual procedure of expanding on angular modes, unfolding the leads, and linearizing around the Fermi points, the Hamiltonian is given by:
\bea
	H &=& H_\R 0+H_\R B, \quad H_\R 0 
	  = -\Ci\sum_{a=1,2} \int_{-\infty}^{\infty}\diff x\;
			\psi^\dagger_a \p^{\phantom{\dagger}}_x \psi^{\phantom{\dagger}}_a\\
	H_\R B &=& \big(
			\gamma^{\phantom{\dagger}}_1\psi_1^\dagger(0)
		  +\gamma^{\phantom{\dagger}}_2\psi_2^\dagger(0) \big) \;d + \hc 
\\%		  		\XXX{\\
&& \hspace*{-0.5cm}+\;
		U\; \big(:\!\psi_1^\dagger\psi^{\phantom{\dagger}}_1\!\!:\!\!(0) 
				  +:\!\psi_2^\dagger\psi^{\phantom{\dagger}}_2\!\!:\!\!(0)\big)\, 
			\big(d^\dagger d-{\textstyle \frac{1}{2}} \big) + \epsilon_d\;d^\dagger d,\nonumber		
	%		}
\eea
where the baths are described by two 1D right-moving fields $\psi_a(x)$. 
The impurity/leads coupling involves  
  the tunneling amplitudes that we parametrize as:
\be
	\gamma_1+\Ci \gamma_2=\gamma\sqrt{2}\;e^{\Ci\Gamma/2}.
\ee
In the following, we consider the (particle-hole symmetric) resonant case with  impurity  onsite energy $\epsilon_d=0$.  

The IRLM bears a duality symmetry exchanging large and small $U$'s \cite{Schiller07}. For an intermediate value of $U$ (of order unity), it is self-dual, and enjoys an additional, hidden, SU(2) symmetry that mix the two wires, with generators $\vec J=\frac{1}{2}\psi^\dagger_a\vec\sigma^{\phantom{\dagger}}_{ab}\psi^{\phantom{\dagger}}_b$ ($\sigma$'s are Pauli matrices) ;
as was shown in Ref.\onlinecite{Boulat08}, as a consequence, the out-of-equilibrium self-dual IRLM  (sd-IRLM) bears a description in terms of dressed quasiparticles (qp's), that are the many-body modes diagonalizing the scattering on the impurity.

Let us now sketch the main lines of the derivation. As usual in a scattering approach, one starts by identifying two classes of asymptotic states, incoming and outgoing, that correspond to states coming from the far left towards the impurity, and escaping to the far right thereof, respectively. Those states span Hilbert spaces $\Hin$ and $\Hout$.
%Those are the states of the form $\psi^\dagger_{a_1}(x_1)\psi^\dagger_{a_2}(x_2)\ldots\psi^\dagger_{a_n}(x_n)\ket{0}$ ($\ket{0}$ is the 'absolute' vacuum state annihilated by all $\psi$'s) with $a_i=1,2$ and $x_i\to-\infty$ ($x_i\to+\infty$ respectively), and they  
The effect of the impurity amounts 
to a linear map $\Hin\xrightarrow{\Cal R}\Hout$ that encodes the fate of an asymptotic \emph{in} state prepared 
in the far past
when time-evolved to far future. As soon as there is an interaction, this linear map becomes a complicated many-body object in the electronic basis.
%: if a single particle electronic state is send towards the impurity, interactions on the impurity leads to the excitation of an arbitrary number of particle-hole pairs. 
Integrability of the IRLM \cite{Filyov}
ensures that one can identify a basis for $\Hin$ and $\Hout$ in terms of pseudo-Fock states built out of a finite number of quasiparticle modes 
$A_\alpha^{\ind{in(out)}}(\theta)$, where $\theta$ is a rapidity parametrizing momentum, $p=\frac{m}{2}\,e^\theta$. Those modes diagonalize the map $\Cal R$, in the sense that they cross the impurity without qp production, and that this property extends to any many-qp state \cite{GZ}. The only effect of the impurity is to change the
qp index $\alpha$.  Formally, one has:
\be 
A_\alpha^{\ind{in}}(\theta)=R_{\alpha\beta}(\theta)A_\beta^{\ind{out}}(\theta)
\label{inRout}
\ee 
with $R$ a scattering matrix. For the sd-IRLM, such a basis can
be obtained via bosonization and a mapping to the anisotropic Kondo model (see \cite{Boulat08} for details). The total charge degree of freedom decouples from the problem, and one is left with a single degree of freedom, the charge imbalance between the two wires. The qp's consist of a soliton and an antisoliton $A_\pm$, and two breathers $A_0$ and $A_1$. Importantly, those qp's fall into representations of the aforementioned SU(2) symmetry: $\{A_\pm,A_1\}$ transform as a spin one, while $A_0$ is a singlet. 
The charge imbalance $\hat Q\!=\!\int \diff x\big(\psi^\dagger_1\psi^{\phantom{\dagger}}_1\!-\!\psi^\dagger_2\psi^{\phantom{\dagger}}_2\big) \!=\! 2\int \diff x J^z$
acts diagonally on the modes: $\hat Q \cdot A_\alpha(\theta) \!=\! q_\alpha A_\alpha(\theta)$, 
with $q_\pm\!\! = \!\! \pm 2e$ and $q_{0,1}\!\!=\!\!0$. Introducing the operator 
$A_\alpha^\dagger$ that destroys the qp $A_\alpha$, the charge 
$\hat Q$ bears a simple representation in terms of the modes: 
$\hat Q=\sum_\alpha \int \diff\theta q_\alpha A_\alpha(\theta)A^\dagger_\alpha(\theta)$. 
To complete the description,  non-vanishing elements of the $R$-matrix  are  
$R_{\pm\pm}\!\!=\!\! \qsg-\psg\cos^2\Gamma $, 
$R_{\pm\mp}\!\!=\!\! \psg \sin^2\Gamma$, $R_{1\pm}\!\!=\!\!R_{\pm 1}\!\! =\!\!
  \psg\frac{\sin(2\Gamma) }{\sqrt{2}}$, $R_{11} \!\!=\!\!
\qsg\!+\!\psg\cos(2\Gamma) $,  and $R_{00}$, with
$\psg(\theta)\!=\!\prod_{k=0,\pm 1}
%\frac{1}{ i e^{\theta-{\thetaB}+\frac{\Ci\pi k}{3}} -1}
\frac{-\Ci}{ e^{\theta-{\thetaB}+\frac{\Ci\pi k}{3}} +\Ci}
$ 
and $\qsg(\theta)\!=\!-\Ci e^{3(\theta-\thetaB)}\psg(\theta)$.
 The lead/impurity coupling results in the appearance of an energy scale ${\TB}=\frac{m}{2}\,e^{{\thetaB}}$ marking the crossover between weak and strong hybridization regimes.

Forcing the sd-IRLM out of equilibrium is achieved by imposing different chemical potentials 
$\mu_{1(2)}=\pm\frac{\V}{2}$ on incoming electronic states in wires 1 and 2, i.e. by coupling 
the system to 
$\frac{\V}{2}\hat Q^\inc=\frac{\V}{2}\int_{-\infty}^0 \diff x \big(\psi^\dagger_1\psi^{\phantom{\dagger}}_1-\psi^\dagger_2\psi^{\phantom{\dagger}}_2\big)$. 
In the qp basis, this operator acts diagonally on the modes $A_\alpha^\inc$, and the voltage $\V$ just amounts to a doping 
of \emph{in} modes.
At zero-temperature and positive voltage, the groundstate is obtained by  filling antisoliton states up to a 
voltage-dependent rapidity $A=\ln(2p_\f/m)$ where $p_\f\propto \V$ is akin to a ``Fermi momentum" (because of interactions, the proportionality constant is not one), with distribution function $\rho^{\!\V}_-(\theta)$ that is determined by doing the thermodynamics for the gas of incoming antisolitons. 
\begin{figure}[t]
	\graphicspath{{./fig_Data/}}
	\begin{footnotesize}\input{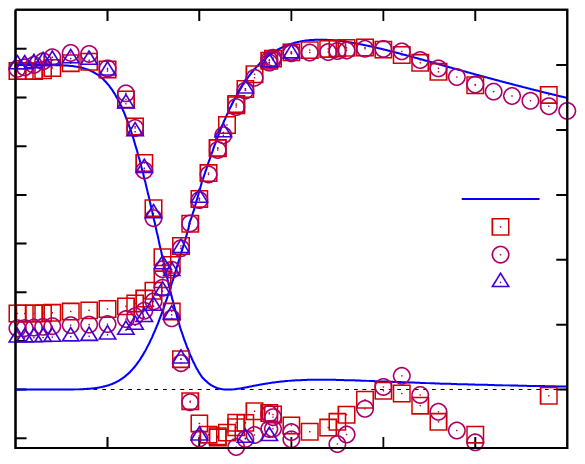}\end{footnotesize}
	\caption{Finite size error of noise. The blue lines represent the analytical result Eq.~\eqref{eq:NoiseResult} with 
	symmetric coupling ($\Gamma\!=\!\frac{\pi}{2}$). The numerical data have been obtained for systems with coupling 
	$J'\!=\!0.4J$ and density-density interaction $U\!=\!2.0 J$ using td-DMRG. The system size varies from $M\!=\!48$ to 
	$M\!=\!72$ lattice sites. The difference of numerical and analytical data in the low voltage regime is proportional to 
	the squared differential conductance $G^2$ and scales linearly with the inverse system size $1/M$.}
\label{fig:raw_data}
\end{figure}

The current operator counts the charge imbalance between \emph{in} and \emph{out} modes, and reads (the factor $\frac{1}{2}$ comes from simple charge counting): $\hat I=\frac{1}{2}(\hat Q^\inc-\hat Q^\out)$, that can be rewritten solely in terms of \emph{in} modes using (\ref{inRout}):
\be
\hat I=\int \diff\theta\;A^\inc_\alpha(\theta) \;\Pi_{\alpha\beta}(\theta)\;A^{\dagger\,\inc}_\beta(\theta)
\label{currOp}
\ee
with $\Pi\!=\!\frac{1}{2}\left( Q \!-\! R^*QR \right)$ and $Q_{\alpha\beta}\!=\!q_\alpha\delta_{\alpha\beta}$. 
Averaging (\ref{currOp}) yields  $I\!=\!\VM{\hat I}\!=\!2\sin^2\Gamma \int_{-\infty}^A \diff \theta \rho^{\!\V}_-(\theta) \Cal T(\theta)$, with $\Cal T\!=\!|\psg|^2$. %=\frac{1}{1+e^{6(\theta-{\thetaB})}}$.  
The zero frequency noise in the steady state, $S_0\!=\! \VM{\hat I^2}-I^2$, is obtained by averaging the square of (\ref{currOp}). 
The terms $\VM{ A^\inc_\alpha(\theta) A^{\dagger\,\inc}_\beta(\theta) A^\inc_{\alpha'}(\theta') A^{\dagger\,\inc}_{\beta'}(\theta')}$ 
that appear have a simple expression at zero temperature \cite{Fendley96}, and the noise reads: $S_0=2\sin^2\Gamma\int_0^{A} \diff\theta\; \rho^{\!\V}_-(\theta)
\big[(1\!+\sin^2\Gamma)\Cal T(\theta)-2\sin^2\Gamma(\Cal T(\theta))^2\big]$.
Simple algebraic manipulations using the scaling form \cite{Boulat08} of the current, $I=\V f(\frac{\V }{{\TB}})$, yield: 
\be
S_0=\cos^2\Gamma\; I + {\textstyle \frac{1}{3}} \sin^2\Gamma\left(I-G\V \right)
\label{eq:NoiseResult}
\ee
with $G\!=\!\frac{\p I}{\p \V }$ the differential conductance. For self-containedness, the scaling function defining the current is: $f(x)\!=\!\frac{G_0\sqrt{\pi}}{2}\bar f(\frac{x}{\alpha})$, 
$\alpha\!=\!\frac{4\sqrt{\pi}\Gamma(2/3)}{\Gamma(1/6)}$,  
$G_0\!=\!G(\V\!\!=\!\!0)\!=\!\frac{e^2\sin^2\Gamma}{h}$ and
$\bar f(x)= \sum_{n\geq 0} \frac{(-1)^n(4n)!}{n!\Gamma(3n+3/2)}x^{6n}$ for $x\!<\!x^*\!=\!\frac{\sqrt{3}}{4^{2/3}}$
while
$\bar f(x)= \sum_{n> 0} \frac{(-1)^{n+1}\Gamma(1+n/4)}{n!\Gamma(3/2-3n/4)}x^{\frac{-3n}{2}}$  for  $x\!>\!x^*$.

In the high voltage limit, the Fano factor $S_0/eI\simeq \frac{1+\cos^2\Gamma}{2}=\frac{\gamma_1^4+\gamma^4_2}{(\gamma_1^2+\gamma_2^2)^2}$ is identical to that of free electrons tunneling through a double barrier 
with tunneling rates proportional to $\gamma_a^2$ \cite{Chen90}. 
Thinking of voltage as a high energy cut-off, this is consistent with the picture of charge carrier being independent electrons \footnote{Nevertheless, even in the ultraviolet limit the interaction manifests itself as a drastic reduction of the number of hybridized current-carrying states
resulting in the suppression of the current $I\sim \V ^{-1/2}$ -- see B. Doyon, Phys. Rev. Lett. 99, 076806 (2007).}.
At low voltage, for a symmetric junction $\Gamma\!=\!\frac{\pi}{2}$, introducing the backscattered current 
$\Ibs\!=\!G_0 \V-I$, one obtains $S_0\simeq 2e\Ibs$, i.e. Poissonian noise for charge $2e$
particles, consistent with antisolitons being the charge carriers at low energy.

\begin{figure}[t]
	\graphicspath{{./fig_Data/}}
	\begin{footnotesize}\input{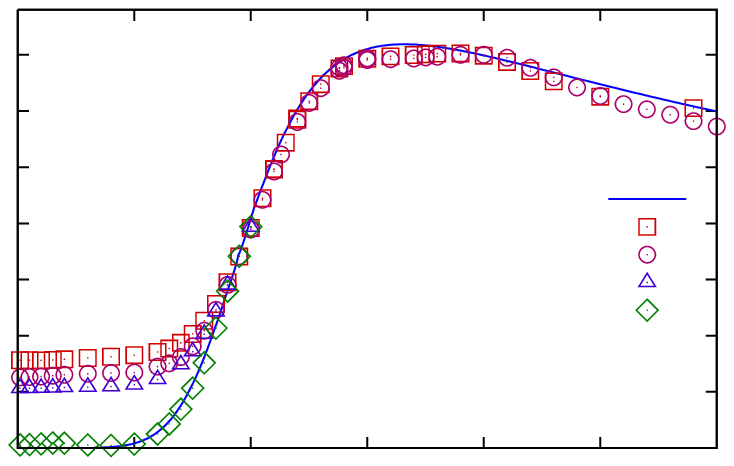}\end{footnotesize}
	\caption{Linear extrapolation of the numerical shot noise data. The linear scaling of the low voltage finite size 
	error is exploited to perform a linear extrapolation $1/M\rightarrow 0$. We find nice agreement of numerical and 
	analytical results.}
\label{fig:extrapolated}
\end{figure}
A completely different approach is based on the numerical simulation of the time evolution of the system. We use the 
td-DMRG \cite{Schmitteckert_PRB2004} to obtain the time dependent current fluctuations 
\begin{flalign}
 S(t,t') =\erwartungswert{\Delta\op I(t) \Delta\op I(t')}_\Psi, ~~&\Delta\op I(t) = \op I(t) - \erwartungswert{\op I(t)}_\Psi, \\
% \end{flalign}
% \begin{flalign}
 \op I(t) = \e^{\Ci \op H t} \op I \e^{-\Ci \op H t}, ~&~ \erwartungswert{\cdot}_\Psi = \bra\Psi \cdot \ket\Psi,
\end{flalign}
for a tight-binding lattice version of the IRLM
\begin{flalign}
\label{eq:Hamiltonian1}
  \op H &= \op H_\R 0 + \op H_\R B, \\
\label{eq:Hamiltonian2}
  \op H_\R 0 &= 
      -J \Big\lbrace \sum_{m=-M_\R L}^{-2} \opcd_{m+1}\opcp_m
      + \sum_{m=1}^{M_\R R-1} \opcd_{m+1}\opcp_m\Big\rbrace+\hc, 
      \nonumber\\
  \op H_\R B &=\!\! \sum_{m=\pm 1}\! \!\Big\lbrace 
		  U \big(\op n_m\!-\!{\textstyle\frac 1 2}\big)\big(\op n_\R d \!-\!{\textstyle\frac 1 2}\big)
		 \!-\!J' (\opcd_{m}\op d\!+\!
		 \hc
		 %\op d^\dagger\opcp_m
		 )\Big\rbrace \nonumber 
		 %\\
      %& \phantom{=} 
       \!+\! V_\R g \op n_\R d,
%\label{eq:Hamiltonian3}
% \nonumber
\end{flalign} 
with the creation, annihilation, and density
 operators in the leads 
($\opcd_j$, $\opcp_j$, $\op n^{\phantom{\dagger}}_j\!=\!\opcd_j \opcp_j$) and on the level 
($\op d^\dagger$, $\op d$, $\op n_\R d\!=\!\op d^\dagger\op d$). 
The total number of sites 
is given by $M\!=\!M_\R L +$ $M_\R R \!+\!1$.
The time dependent fluc\-tua\-tions then allow for the calculation of the noise power spectrum \cite{BlanterButtiker1999} by means of a Fourier transform 
$S(\omega)=4\Re\int_0^\infty \diff t\, \e^{\Ci\omega t}S(t,t')$, where we now restrict ourselves to the low frequency 
limit $\omega\!=\!0$. 
Note that the discrete nature of the leads results in a finite bandwidth $4J$ with cosine dispersion.
We create an initial state by taking the ground state of $\op H\!+\!\V(\op N_\R L\!-\!\op N_\R R)/2$, 
where we added a charge imbalance operator to the system $\op H$
in order to create different fillings in the left and right lead.
We now perform a time evolution with respect to $\op H$ and search for a stationary regime 
close to
the impurity, for details see \cite{Branschaedel_Schmitteckert:X2010}.

Within this approach we have to address the following time scales. 
First, initial oscillations of the current die out on a time scale $t_\R S$ 
which is typically proportional to the inverse of the resonance width \cite{wingreen93,Branschaedel_Schmitteckert:X2010}. 
Second, at transit time $t_\R R \!=\! M/v_\R f$ the 
reflections at the boundaries reach the impurity and we have to stop our simulations.
Third, after choosing a starting point $t' \!\gg\!  t_\R S$ we have to simulate until $t_{\R{max}} \!<\! t_\R R$, 
where $t_{\R{max}}  \!-\! t'$ has to be large enough to allow for a faithful calculation of the noise power spectrum. 
A detailed analysis of the finite size correction  in 
 the non-interacting case is provided in \cite{previous}.
% 
% The details of the numerical simulation based on the td-DMRG, of the preparation of the initial state and of the 
% finite size effects seen in the current expectation value have been discussed extensively in the past couple of years 
% \cite{xyz} while the extraction of shot noise from time evolution simulations including the subtraction of finite size 
% effects was investigated only recently \cite{xyz} for the case of the non-interacting RLM. There, the density-density 
% interaction was sent to $U=0$ allowing for the representation in a single particle picture and the application of 
% exact diagonalisation algorithms to treat the problem. Since the comparison of numerical data and analytical result 
% showed very nice agreement, it seems justified to apply the same approach to the IRLM. 

Here, we set $U=2.0 J$, corresponding to the 
sd-IRLM
as discussed in \cite{Boulat08}, and the coupling to 
$J'=0.4J$, while we operate in the resonant tunneling regime $V_\R g=0$. The total number of lattice sites varies from 
$M=48$ to $72$ lattice sites, with $M_\R R=M_\R L+1$. Different other setups have been considered, including the 
effective enlargement of the system using damped boundary conditions, which will not be presented in this work. For 
the numerical simulation within the DMRG projection scheme we set an upper bound to the dimension of the Hilbert space 
for each DMRG block to $N_\text{cut}=4000$ states. 

As a first result we compare the numerical data for different system sizes to the analytical result in 
Fig.~\ref{fig:raw_data}, where we show zero-frequency shot noise as well as the finite size error of the numerical 
data, rescaled by the system size. In the low frequency limit, strong finite size effects have to be expected, that 
get mostly pronounced for small values of the voltage \cite{previous}. Since the rescaled finite size error happens to 
collapse on a single curve in the low voltage regime, the numerical data can be linearly extrapolated to infinite 
system size in order to obtain results for the thermodynamic limit. Also we verify an analytical estimate for the 
finite size error $S_\R{num.}-S_\R{analyt.}\propto G^2/M$ with 
$G$ the differential conductance 
that has been given for the free case in \cite{previous}.
The strong deviations in the high voltage regime from this relation may be traced back to different sources: the 
approximative td-DMRG scheme introduces a cutoff error that gets especially pronounced for values of the voltage of the 
order of the bandwidth. Furthermore, to keep the numerical simulation feasible, one has to resort so small systems 
introducing finite size effects beyond the linear scaling.

\begin{figure}[t]
	\graphicspath{{./fig_Data/}}
	\begin{footnotesize}\input{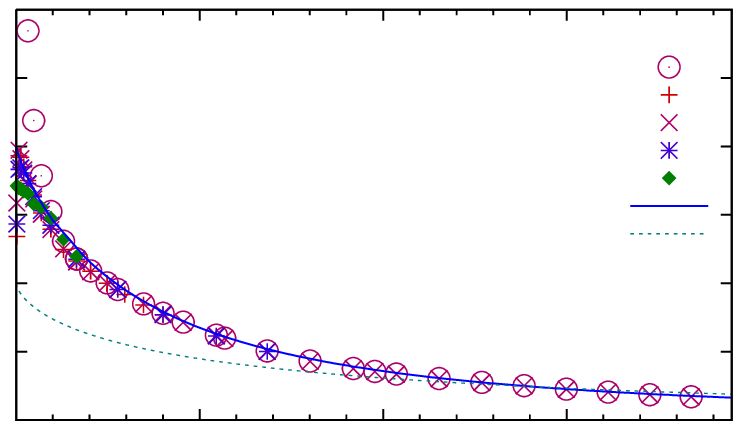}\end{footnotesize}
	\caption{Back scattering Fano factor as a function of the back 
	scattered current. The numerical data points have
	been obtained using the numerical shot noise data divided by the analytical back 
	scattered current. The finite 
	size error of the numerical results for
	 shot noise leads to a diverging Fano factor. The situation 
	improves for the linearly extrapolated data, while we find a nice agreement of the analytical result with 
	the $G^2$-corrected data. For comparison we 
	show the Fano factor in the non-interacting case.}
\label{fig:fano}
\end{figure}

Nevertheless, the numerical results shown in Fig.~\ref{fig:extrapolated}, where we obtained data for the low voltage 
regime using linear extrapolation, show very nice agreement with the analytical results given by 
Eq.~\eqref{eq:NoiseResult} with symmetric coupling, $\Gamma\!=\!\frac{\pi}{2}$. 
%%On the one hand this confirms the validity of the 
%%analytical calculation, while, on the other hand, it extends the applicability of the prescription introduced in 
%%\cite{previous} to strongly correlated systems. 
The back scattering Fano factor $\Fbs\!=\!S/\Ibs$, 
%$\Ibs\!=\!G_0 \V \!-\! I$, 
can also be obtained from the numerical data, Fig.~\ref{fig:fano}, where we use the analytical result for the current \cite{Boulat08}. It fits nicely with the analytical result for $\Fbs$ as long as finite size effects can be neglected. 
However, the finite size offset at $\Ibs\rightarrow 0$ leads to a strongly diverging Fano factor. %, when no finite size corrections are applied. 
In contrast, $\Fbs$ remains finite even for very small values of $\Ibs$, when obtained from the linearly extrapolated shot noise data. The deviations from the analytical result at small $\Ibs$ can be traced back to small absolute errors that get blown up in the limit $\Ibs\rightarrow 0$. The very nice agreement of analytical result and $G^2$-corrected data, even in the regime of very small $\Ibs$, indicates that increasing the system size and adding more data points to the extrapolation procedure should improve the extrapolated result.

% \begin{table}[t]
%   \begin{tabular}[t]{C{0.2\columnwidth}C{0.2\columnwidth}C{0.2\columnwidth}|C{0.2\columnwidth}c}
% 	   {$M$} &    {$J'/J$} &    {$V_\R g$} &    {$N_\text{cut}$}  \\ \hline
% 	   {$48$}&    {$0.4$}  &    {$0$}      &    {max. $4000$}     \\ 
% 	   $60$  &     $0.4$   &     $0$       &    max. $4000$       \\
% 	   $80$  &     $0.4$   &     $0$       &    max. $4000$
%   \end{tabular}
%   \caption{Parameters.}
% \end{table} 

In summary we have provided two different methods, an 
analytical 
%field theoretical 
approach within the framework of
the thermodynamic Bethe ansatz, and time dependent DMRG simulations on the lattice to obtain the noise
correlations in the IRLM. Both methods show excellent agreement and provide benchmark results for other methods. 
Most strikingly our results show a strong enhancement of the back scattered Fano factor due to interaction effects.

On the conceptual side, we believe our result further establishes the reliability of the Bethe ansatz approach to transport pioneered in \cite{FLS95}. One of the objections to this approach  sometimes raised is that it relies on the theory of excitations over the vacuum, and thus deals with fundamental objects - quasiparticles - which are not simply related to the bare electrons. In the present case, these objects are  solitons, of charge $|q_\pm|=|2e|$, which are made of combinations of particle hole pairs mixing the two wires. The results in the low energy limit give these quasiparticles their physical reality: they are the objects that tunnel in a Poissonian way at low voltage, and the  Fano factor is  a direct measure of their charge.

\acknowledgments                                                                                                                                                          
%{\bf Acknowledgments: }
The DMRG calculations have been performed on HP XC4000 at Steinbuch Center for Computing (SCC) Karlsruhe under project RT-DMRG.        
We acknowledge the support by the DFG Center for Functional Nanostructures (CFN), project B2.10. 

\bibliographystyle{unsrt}

\end{document}